# CONCEPT OF PHOTONIC HOOK SCALPEL GENERATED BY SHAPED FIBER TIP WITH ASYMMETRIC RADIATION

Igor V. Minin[1,2,*], Oleg V. Minin[1,2], Yan-Yu Liu[3], Cheng-Yang Liu[3,**]

[1]Tomsk Polytechnic University, Tomsk, Russia
[2]National Research Tomsk State University, Tomsk, Russia
[3]Department of Biomedical Engineering, National Yang-Ming University, Taipei City, Taiwan

*E-mail: prof.minin@gmail.com
**E-mail: cyliu66@ym.edu.tw

Structured light have made deep impacts on modern biotechnology and clinical practice, with numerous optical systems and lasers currently being used in medicine to treat disease. We demonstrate a new concept of fiber-based optical hook scalpel. The subwavelength photonic hook is obtained in the vicinity of a shaped fiber tip with asymmetric radiation. A 1550 nm continuous-wave source, commonly used for medical imaging, has been required. Photonic hook with a lateral feature size less than the half-wavelength is achieved using a hemispherical shaped fiber tip with metallic mask. This breakthrough is carried out in ambient air by using a 4-μm-diameter fiber with a shaped tip. A good correlation is observed between the computed intensity distribution of photonic hook and the tip sizes. Photonic hook generated with a shaped fiber tip, easier to manipulate, shows far-reaching benefits for potential applications such as ophthalmic laser surgery, super-resolution microscopy, photolithography, and material processing.

Introduction

The use of light in medicine began thousands of years ago. The first laser light interaction with tissue was, probably, studied experimentally by Milton M. Zaret *et al* in 1961, who exanimated photocoagulation in the retina [1]. Three years later C. Kumar N. Patel invented the carbon dioxide gas laser in 1964 [2]. Later this gas laser was used for the first time to perform soft tissue surgery. L. Goldman and R.G. Wilson studied the destruction of skin lesions in 1964 [3]. Today one of the key elements of this medical device is a laser scalpel connected with an optical fiber to add of flexibility [4]. Different focusing devices are used on a fiber end to improve the light radiation exposure efficiency [5-7]. Focusing of multimodal light beams may be used in ultraprecise laser procedures [8] in the brain and eye with help of chains of dielectric microspheres assembled inside the cores of hollow waveguides [9,10]. It was shown that near a shaped optical fiber photonic jets (PJs) can be obtained at wavelength of 1064 nm [11,12]. Such structures [8-11] can focus laser beam into tissue in 'contact mode' with the tip of the optical scalpel and can ablate the tissue. Moreover, such laser scalpels were used for dissecting and removing fibrotic membranes from the retina surface [13]. In ophthalmic laser surgery, for example, the micro-hook needles are used to remove deposits from the retina for proliferative diabetic retinopathy [14]. The metal micro-hook scalpel has been used in neurosurgery for a long time due to its little contact with adjacent tissues [15].

The idea of application of recently discovered [16] a photonic hook (PH) effect [17,18], based on dielectric Janus particle, as the ultraprecise laser scalpel for the first time was offered in [19]. It has recently been shown, that a PH in free space can also be obtained by asymmetric illumination of symmetric microparticle [20,21].

In this paper, we propose a fiber-based laser hook scalpel concept which the curvature and length of need to be carefully balanced. In this study, we report on the formation of a curved optical hook created by fiber with mesoscale hemispherical particle with partially blocking fiber tip by an amplitude mask height $h$. The key parameters of the PH such as lateral full width at half maximum (FWHM), curvature, and maximal field intensity depending on the variation of the amplitude mask height are studied systematically. The wavelength of the laser beam is 1550 nm. The materials of the fiber and hemispherical tip are pure silica with refractive index of 1.44 [22]. The diameter of multimode optical fiber is 4 μm.

Simulation model

The objective lens is used to focus the laser beam into the fiber core. The metal mask is made of aluminum. In this study, we conduct finite-difference time-domain (FDTD) simulations [23] to analyze the effect on PH structure, illustrated in Figure 1. The simulation domain is set to be 3 μm × 10 μm, the grid size of the FDTD-mesh is 30 nm, and external boundary conditions are the perfectly matched layer for diminishing the effect of reflected lightwave at the boundary.

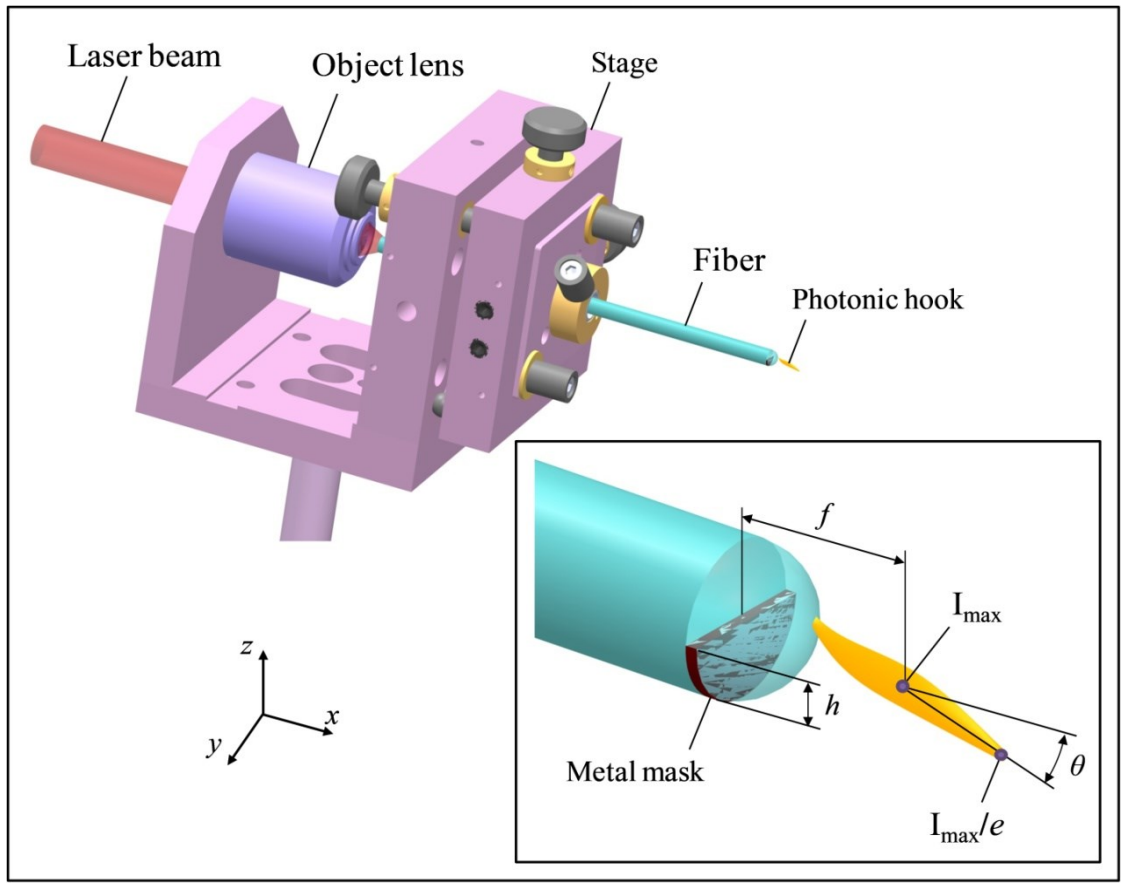

Figure 1. Schematic diagram of the fiber tip with metallic mask. The insert indicates the definition of the photonic hook.

Results of simulations

To illustrate the effect of the irradiating beamwidth on the PH formation, Figure 2 shows the power flow patterns of the PHs formed by fiber with hemispherical particle tip with different mask heights. The width of the irradiating beam is regulated by the height of the amplitude mask. It can be clearly seen that the PH curvature and length changes as the mask height increases. However, the PH position moves inside the fiber tip when the height of the amplitude mask is more than $0.35d$.

Following to [20-21] the physics of the curved PNJ formation can be explained as follows. The refractive index of fiber and width of illumination determines the angle of refraction on an interface via the generalized Snell's law [24]. Part of the illuminating beam inside the fiber and determined by the mask height, is diffracted first on the metall mask. Than the light beam inside the fiber is refracted when it exits from the shadow surface of the fiber tip (see Fig. 1). If the width of the illuminating beam is less than the fiber diameter, the components of the wave vector K∥ do not cancel each other inside the fiber through the local destructive interference [20-21]. The wave vector K∥ is relative to the axis of symmetry of the fiber, which creates the PH curvature profile. On the other hand, the components of the wave vector K⊥ determine the length of the PH along the propagation direction. Therefore, the local optical fields interfere inside the fiber and then the PH can be generated outside the fiber tip depending on the width of the illuminating beam.

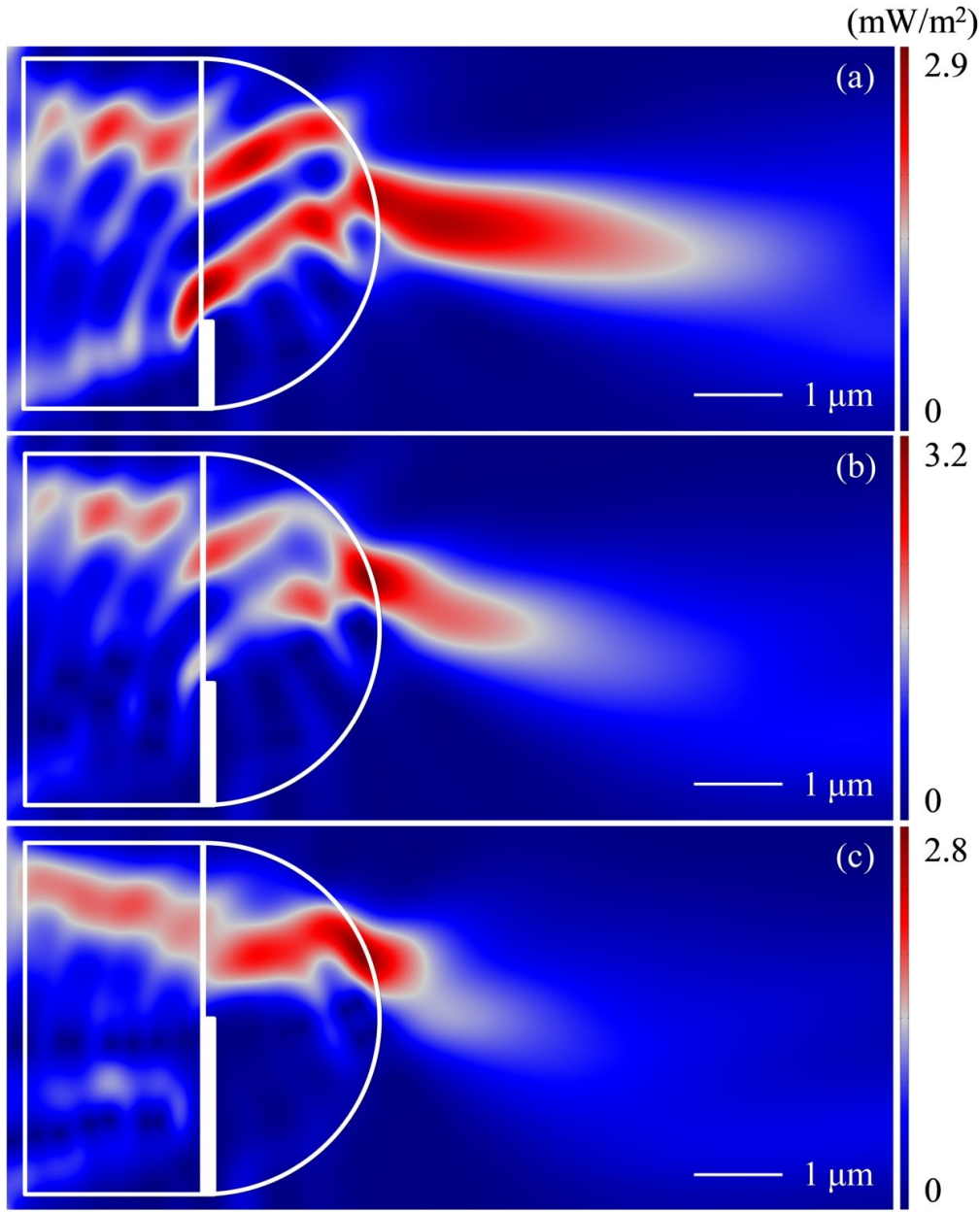

Figure 2. Power flow patterns of the photonic hooks formed by fiber tip with metallic mask at (a) $h = 0.25d$, (b) $h = 0.35d$, and (c) $h = 0.5d$, excited by the fundamental mode of the fiber.

The difference between the intensity distributions generated at different mask height $h$ values can be compared by the field intensity profiles. Figure 3 shows the normalized field intensity profiles of the PHs along the transverse direction, measured at maximal peak intensity plane. It can be seen that FWHM of the PH at $h = 0.25d$ is $0.55\lambda$. For other values of the mask height it could be noted that maximal field intensity peak is located inside the hemispherical tip of the fined with refractive index $n = 1.44$. The FWHMs at $h = 0.35d$ and $h = 0.5d$ are $0.42\lambda$ and $0.4\lambda$, respectively.

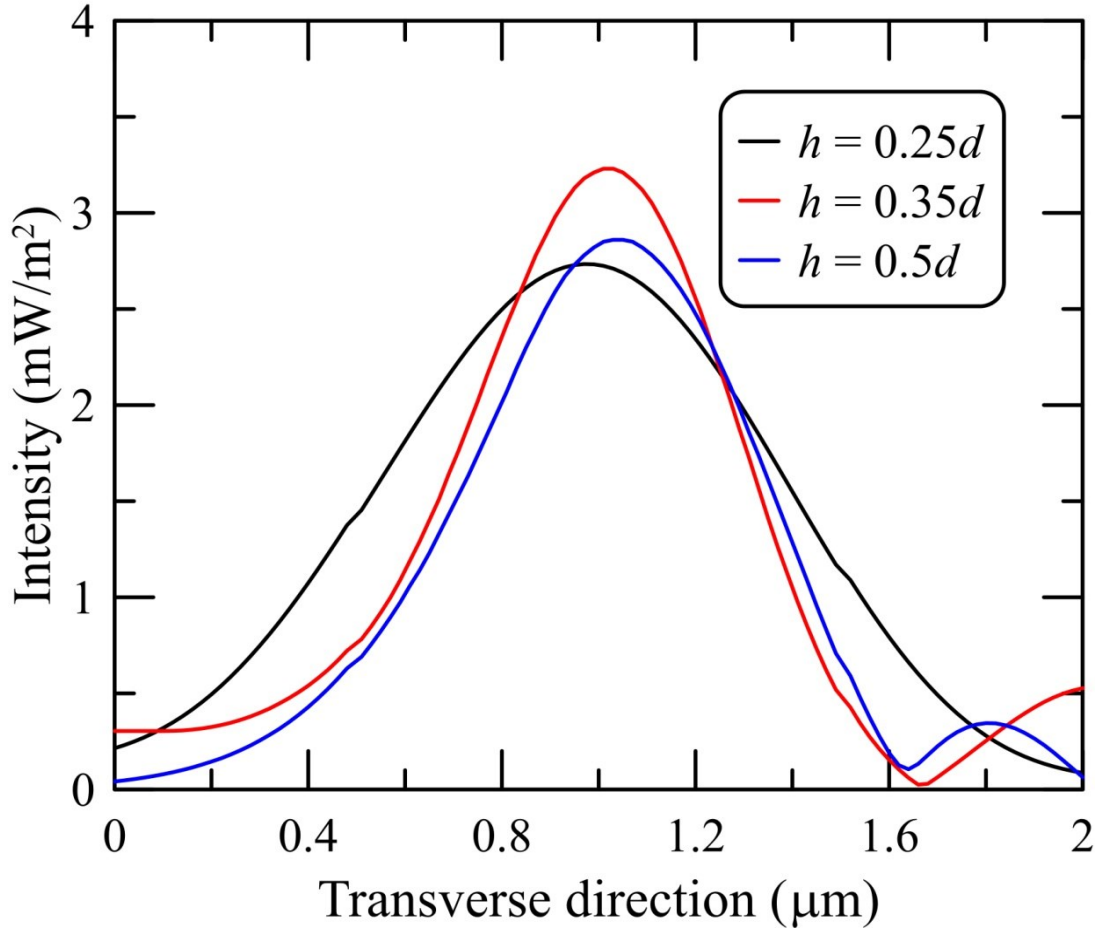

Figure 3. Intensity profiles of the photonic hooks for fiber tip along transverse direction.

The photonic hooks key parameters (focal length, FWHM, and tilt angle) as a function of the mask height, which define the curvature of the PH [17-18], are shown in Figure 4. For evaluating the tilt angle and the FWHM of the PHs [25], we find the inflection point (Imax) in the simulation images. The FWHM is defined as the double distance perpendicular to the propagation direction between Imax and the half-maximum point. If the whole PH is generated outside the fiber tip, the end and start points are located at $1/e$ Imax on the left and right sides of Imax along the PH. If the PH distributes across the fiber tip boundary, the start point is the intersection point of the PH and the fiber tip boundary. The left (start-inflection points) and right (end-inflection points) PH arms relative to the direction of radiation incidence are obtained as shown in Fig. 1. The tilt angle $\theta$ is identified as the supplementary angle between the left and the right PH arms.

The dash line in Fig. 4(a) corresponds to the boundary of hemispherical fiber tip. It can be clearly seen that with mask height less than *h≦0.32d* the maximal field intensity position along the PH is out of fiber tip. The diffraction limited PH (FWHM $\leqq \lambda/2$), equal to FWHM=0.45$\lambda$, is observed at mask height *h=0.32d* with focal length of F=1.3$\lambda$ and tilt angle about 25 degree. Also one can see the mask height has a significant influence on the curvature of the PH, as expected.

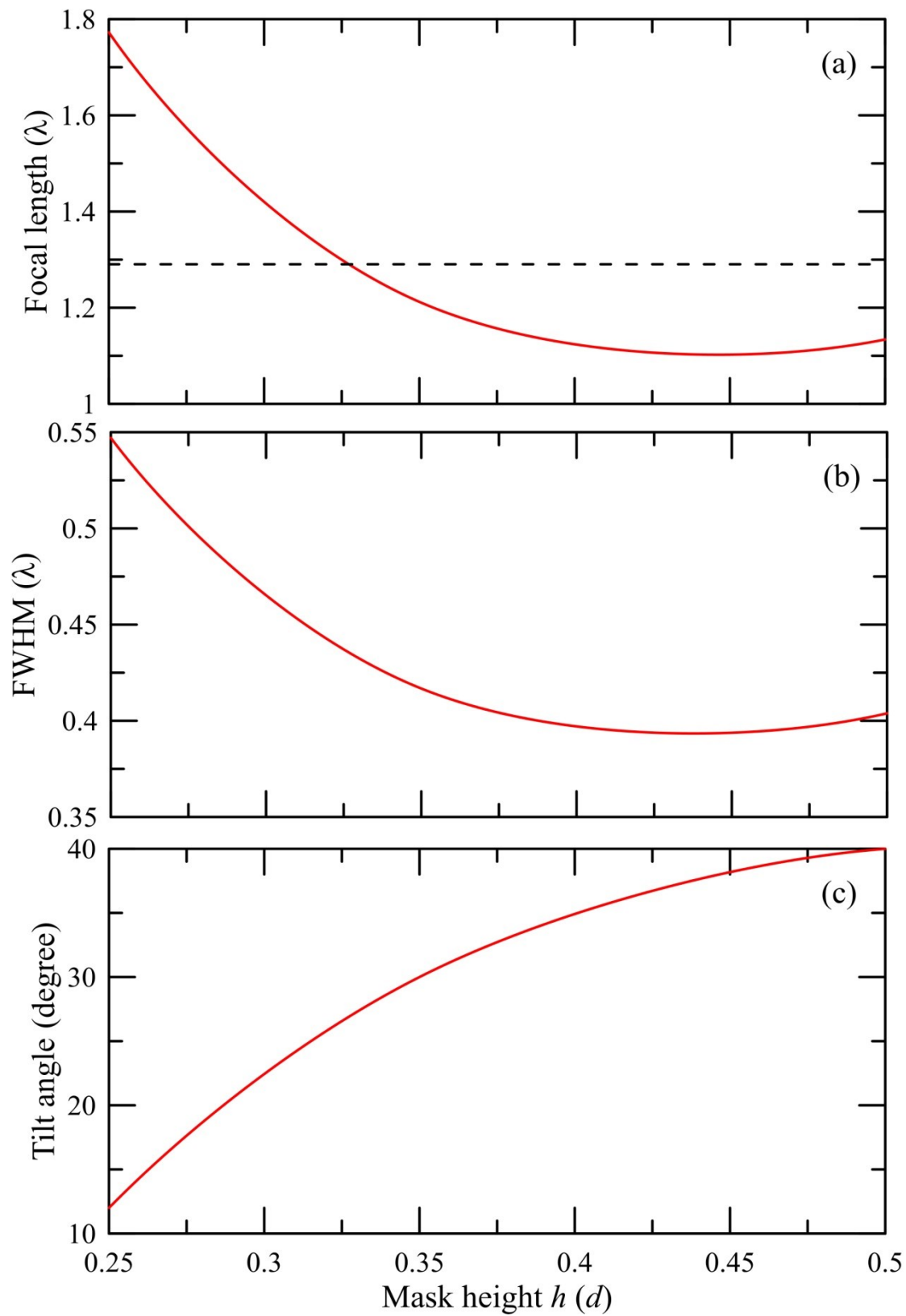

Figure 4. Key parameters as a function of the mask height for photonic hooks: (a) focal length, (b) FWHM, and (c) tilt angle. The dash line in Fig. 4(a) is the boundary of hemispherical tip.

Conclusions

In conclusion, we have described the concept of optical hook fiber scalpel based on photonic hook phenomenon. Different laser scalpel configurations with varying metal mask dimension were successfully simulated demonstrating the concept of optical fiber-based hook scalpel. We have shown that the spatial shape and curvature of the PH with FWHM below the diffraction limit can be simply tailored by varying the metal mast inserting into the fiber tip. The correct combination of the illuminating beamwidth vs mask height enables the efficient beam shaping and directionality of the PH launching. The orientation of this PH with a tissue contact strongly determines the heated area. Thus, fibers based photonic hook enable to optimize tissue cutting modes due to controllable curved shape of the field localization area. Using the FDTD technique, we have demonstrated the key characteristics of the PH optical scalpel formed in the vicinity of the hemispherical fiber tip with metallic mask. In practical applications, a gel-like medium may be used to mimic the optical properties of a biological tissue [26,27]. Photonic hook is likely to make a significant contribution is the medical field and more precisely the field of surgery with a minimally traumatic. Sapphire waveguides are also of potential interest, but their drawback are a high refractive index and fragility of the material [28,29].

The glancing angle deposition in sputtering system can be used to produce the fiber tip with metal mask [20]. The refractive index of a metal mask is varied by using different metal targets during deposition. We are currently building glancing angle deposition system for coating metal mask on the flat surface of the fiber tip. The further investigations of such instruments should be done. Moreover, the results of this work may be of interest for surface texturing [31-33] and nanostructuring, biosensing [34], fiber-based sub-wavelength optical micromanipulation [35-37], improvement of data storage, engraving, micro-welding, to name a few. This will be investigated.

Disclosures and Contributions. The authors have no relevant financial interests in this article and no potential conflicts of interest to disclose. Contributions: I.M. and O.M initiated the ideas of this work and directed the project, wrote the manuscript. Y.Y.L. and C.Y.L. made simulations, I.M., O.M., and C.Y.L. analyzed results. All authors contributed to discussions and commented on the manuscript.

Acknowledgments

This study is funded by Ministry of Science and Technology of Taiwan under Grant Nos. MOST 108-2221-E-010-012-MY3 and MOST 109-2923-E-010-001-MY2, and Yen Tjing Ling Medical Foundation under Grant No. CI-109-24, and is partially supported by Russian Foundation for Basic